\documentclass[10pt,twocolumn,twoside]{osajnl}
\usepackage{csquotes}
\usepackage{amsmath}

\usepackage[utf8]{inputenc}
\usepackage[english]{babel}

\journal{ol} 

\setboolean{shortarticle}{true} 

\title{A large scale passive laser gyroscope for Earth rotation sensing}

\author[1]{K. Liu}
\author[1]{ F. L. Zhang}
\author[1, $\dagger$]{Z. Y. Li}
\author[1]{X. H. Feng}
\author[1]{K. Li}
\author[1]{Z. H. Lu}
\author[2]{K. U. Schreiber}
\author[1, 3]{J. Luo}
\author[1, $*$]{J. Zhang}

\affil[1]{MOE Key Laboratory of Fundamental Physical Quantities Measurements \&
Hubei Key Laboratory of Gravitation and Quantum Physics\\
PGMF and School of Physics, Huazhong University of Science and Technology, Wuhan 430074, P. R. China}
\affil[2]{Technical University of Munich, Forschungseinrichtung Satellitengeod\"asie, Geod\"atisches Observatorium Wettzell, 93444 Bad K\"otzting, Germany}
\affil[3]{TianQin Research Center for Gravitational Physics and School of Physics and Astronomy, Sun Yat-sen University (Zhuhai Campus), Zhuhai 519082, P. R. China}

\affil[$\dagger$]{Email: zongyang\underline{~~}li@hust.edu.cn}
\affil[$*$]{Corresponding author: jie.zhang@mail.hust.edu.cn}

\dates{Compiled \today}

\ociscodes{(120.3940) Metrology; (120.5790) Sagnac effect; (140.2020) Diode lasers; (140.3370) Laser gyroscopes; (140.3410) Laser resonators.}

\doi{\url{http://dx.doi.org/10.1364/XX.XX.XXXXXX}}

\begin{abstract}
Earth rotation sensing has many applications in different disciplines, such as for the monitoring of ground motions, the establishment of UT1 and the test of the relativistic Lense-Thirring effect on the ground. We report the development of a $\bf1~m\times1~m$ heterolithic passive resonant gyroscope (PRG). By locking a pair of laser beams to adjacent modes of the square ring cavity in the clockwise and counter-clockwise directions, we achieve a rotation resolution of about $\bf2\times10^{-9}$~rad/s at an integration time of 1000~s. The sensitivity of the PRG for rotations reaches a level of $\bf2\times10^{-9}~rad/s/\rm \sqrt Hz$ in the 5-100~Hz region, currently limited by the detection noise, residual amplitude modulation and the mechanical instability of the cavity. Our initial results improve the reported rotation sensitivity of the PRGs and indicate that PRGs have a great potential for high-resolution Earth rotation sensing.
\end{abstract}

\setboolean{displaycopyright}{true}

\begin{document}

\maketitle
Many large scale ring laser gyroscopes (RLG) have been built over the last three decades with extremely high sensitivity and stability \cite{RSI:Schreiber:2013}. These large ring lasers along with the developed fiber optic gyros have found applications in different fields like geodesy \cite{RSI:Schreiber:2013}, seismology \cite{RSI:Schreiber:2013,           SCI:Hand:2017, JS:Velikoseltsev:2012, OL:Clivati:2013} and fundamental physics tests \cite{CRP:Virgilio:2014, PRD:Bosi:2011, EPJP:Tartaglia:2017}. The monolithic G-ring experimentally resolves a rotation rate of $\rm 3.5\times10^{-13}$~rad/s over 1000~s and detects the Chandler and the annual wobble of the Earth \cite{OL:Schreiber:2013, PRL:Schreiber:2011}. The successful operation of two really large heterolithic RLGs, namely UG-1 and UG-2 with enclosed areas of 367~$\rm m^2$ and 834~$\rm m^2$, demonstrate the feasibility of constructing giant rotation sensors \cite{AO:Dunn:2002, JAP:Hurst:2009}. Laser gyros are also used to improve the seismic isolation system for ground-based gravitational-wave antennas \cite{CQG:Virgilio:2010, CQG:Korth:2016}. Moreover, an ambitious plan named GINGER has been proposed, focusing on measurement of the Lense-Thirring effect in a terrestrial laboratory \cite{EPJP:Virgilio:2017}. Highly sensitive interferometric fiber optical gyroscopes (I-FOG) are developed mainly for applications in navigation and platform control, with a sensitivity in the order of $10^{-8}\sim10^{-9}$~rad/s/$\rm \sqrt{Hz}$. Such rotational sensors with low self-noise are also suitable for applications in geosciences \cite{ICINS:Korkishko:2013, SRL:Bernauer:2018, Blueseis_website}.

The measurement principle of RLGs is based on the Sagnac effect: the counter propagating beams in a ring cavity will see a different round trip time if the ring cavity is rotating in the optical plane. This effect was first described by Sagnac in 1913 \cite{CRASP:Sagnac:1913}. In 1963, Macek and Davis utilized a ring cavity that contains He-Ne gas to lase at the 1.153~$\mu$m line, which demonstrated rotation sensing \cite{APL:Macek:1963}. This is considered to be the first active RLG. A passive gyro was successfully realized by Ezekiel and Balsamo in 1977 \cite{APL:Ezekiel:1977}, where an external laser is split into two beams and locked to a passive ring cavity in the clockwise (CW) and counter-clockwise (CCW) direction. The rotation rate is determined by measuring the resonance frequency difference of the ring cavity in the opposite directions. A modern PRG, used as a tilt sensor, recently reported a sensitivity of $10^{-8}$ ~rad/s/$\rm \sqrt Hz$ above 0.5~Hz \cite{CQG:Korth:2016}.

The operation principles of an active RLG and a PRG are the same: the resonant frequency difference of the ring cavity in the opposite directions is proportional to the rotation rate of the cavity frame itself,  and can be written as:
\begin{equation}
f_s=\vec{K}\cdot\vec{\Omega},
\end{equation}
where $ f_s $ is the resonant frequency difference, called Sagnac frequency, $ \vec{\Omega} $ is the rotation rate vector, and $ \vec{K}=4\vec{A}/\lambda P$ is the scale factor with $\vec{A} $ the area vector enclosed by the cavity, $ P $ the cavity perimeter and $ \lambda $ the laser wavelength. The main difference between an active RLG and a PRG is that the former has a lasing medium inside the cavity, while the latter utilizes an external injection laser. Both of them have their own advantages and disadvantages. However, in the end they must have the same ultimate sensitivity, which is limited by the cavity loss \cite{PRA:Abramovici:1986, PRA:Gea:1987}. So far the performance of a PRG is still below that of a RLG \cite{CQG:Korth:2016, APL:Ezekiel:1977, OL:Sanders:1981, SPIE:Shaw:1984, JGCD:Lorenz:1988}, so it is worth to revisit these experiments, to look for other limitations and try to improve the technologies further.

In this letter, we report on a heterolithic $\rm1~m\times 1~m$ PRG utilizing a diode laser locked to adjacent longitudinal modes of a ring cavity with a detection noise limited performance. This is a prototype of the PRG that is under development with the initial goal to support the space-borne gravitational wave detector TianQin \cite{CQG:Luo:2016}. As a scientific goal an improved PRG structured on the Earth can contribute to the link of the celestial to the terrestrial reference frame with high-resolution Earth rotation measurement in addition to the current space geodetic techniques \cite{RSI:Schreiber:2013, JOG:Nilsson:2012}.

\begin{figure}[htbp]
    \centering
    \includegraphics[width=\linewidth]{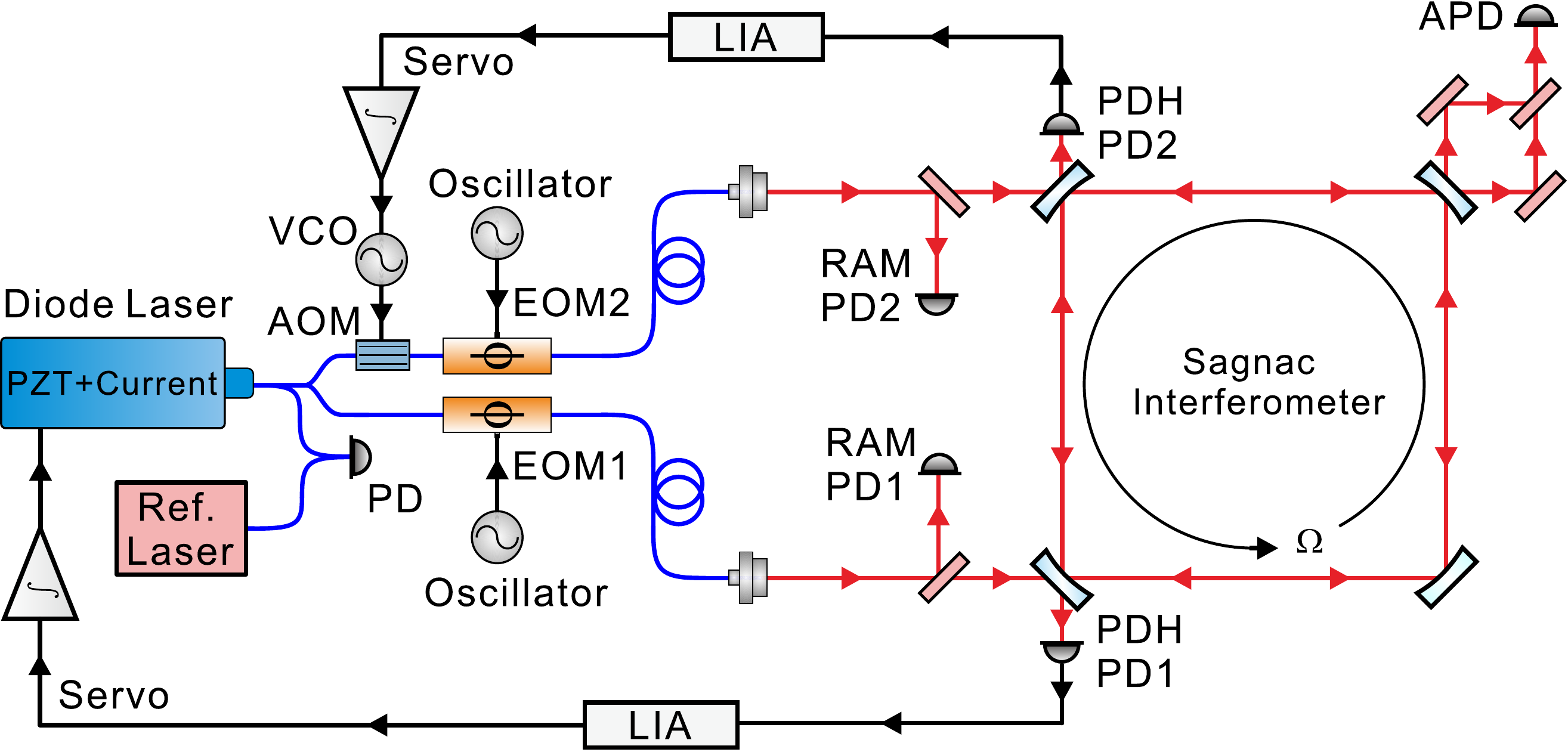}
    \caption{Experimental scheme of the PRG. AOM, acoustic-optic modulator; EOM, electro-optic modulator; RAM PD, photodiode for residual amplitude modulation detection; PDH PD, photodiode for PDH locking; LIA, lock-in amplifier; VCO, voltage controlled oscillator; APD, avalanche photodiode; Ref. Laser, an ultra-stable laser as a reference to diagnose the ring cavity displacement noise.}
    \label{setup}
\end{figure}

The experimental setup is shown in Fig. \ref{setup}. The square ring cavity is made from four super mirrors with a reflectivity of $99.999\%$ and 3~m radius of curvature. We achieve a cavity finesse of about $141,000$ with a measured ring-down time of about 300~$\mu$s. It indicates a high $Q$ factor of $5.3\times 10^{11}$. In order to isolate the air flow, the ring cavity is placed in vacuum with a pressure of $1\times10^{-6}$~Pa. The vacuum system consists of four corner chambers, some connection pipes and bellows. The corner chambers are anchored to a granite platform to enhance the dimensional stability and the bellows are used to reduce the stress caused for example by thermal expansion coefficient mismatching. The four super mirrors on stainless steel mirror mounts are rigidly attached to the chamber bases. The granite platform sits on a frame of six steel legs and is housed in a cave laboratory, which has low seismic noise and small temperature fluctuations.

As shown in Fig. \ref{setup}, the output from a 1064~nm diode laser is split into two branches, each of which is phase modulated by an electro-optic modulator. The two modulated laser beams are mode matched, then coupled into the ring cavity, one in the CCW and the other in the CW direction. Each beam is locked to the ring cavity with the Pound-Drever-Hall (PDH) technique \cite{APB:Drever:1983}. The two locking loops are named the primary and the secondary loop, respectively. The secondary loop is frequency shifted by an acoustic-optic modulator (AOM) to maintain a frequency lock with an offset of nominally one cavity free spectral range (FSR). In the PDH locking scheme, residual amplitude modulation (RAM) can ruin the stability of the locking zero baseline \cite{APB:Shi:2018}. We use two PDs to monitor this effect and then actively stabilize the RAM fluctuations. Finally the two leak-out beams of the Sagnac interferometer are superimposed on an avalanche photodiode (APD) that is used to sense the laser frequency difference. 

In a PRG, a rigid cavity lock is essential to achieve a good sensitivity. To obtain a high signal-noise ratio of the error signal, we take great care of the mode matching of the TEM$_{00}$ mode to the ring cavity. Since the cavity is made of four identical curved mirrors, we measure the Gaussian beam parameters of the leaked light from the cavity by a beam profiler. The measured data is then used as a guidance for the mode matching telescope design. We then achieve a mode overlapping ratio of 70\% for both locking loops. 

Since the linewidth of the free-running diode laser is about 500~kHz, we implement a two-branch feedback control to the diode laser, in order to lock the laser to the ring cavity in the primary loop. The fast feedback branch is set by controlling the current and has a bandwidth of 1.5~MHz. The slow feedback branch uses a piezo inside the laser head with a locking bandwidth of around 5~kHz. It increases the low frequency gain to ensure that the laser follows the cavity length change. Ultimately, a servo feedback gain of 140~dB around 1~Hz has been obtained via the piezo feedback branch. In the secondary loop, the AOM is used as the actuator to accomplish the locking loop by driving a voltage controlled oscillator. The driving frequency is denoted as $f_{AOM}$. In our case,  $f_{AOM}$ is approximately 75~MHz to match the FSR of the square ring cavity. In this way, locking perturbations caused by the back-scattered light can be significantly reduced. 

\begin{figure}[htbp]
    \centering
    \includegraphics[width=\linewidth]{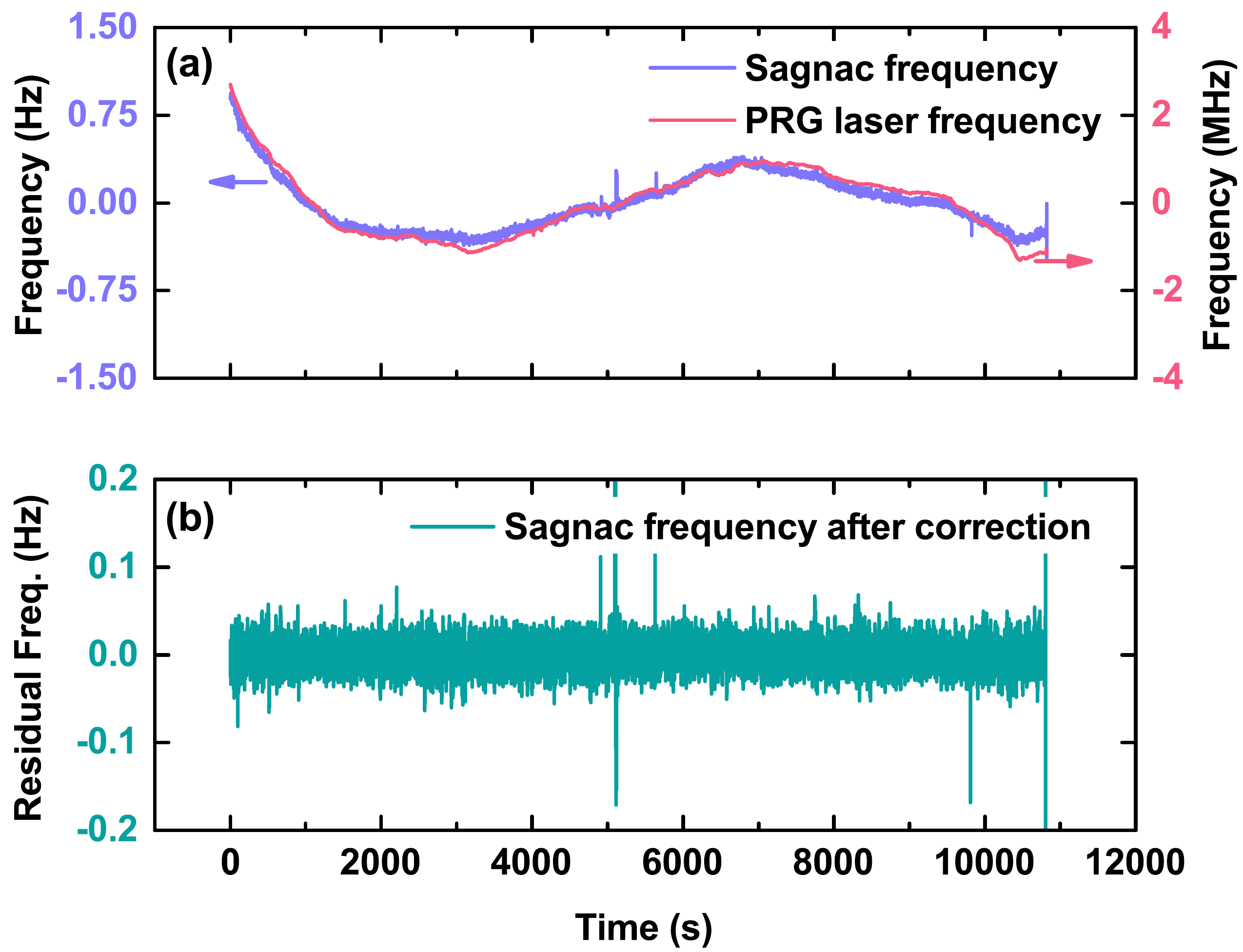}
    \caption{(a) The detected Sagnac frequency is shown with a blue curve, and the resonant PRG laser frequency variation of the primary loop measured with an ultra-stable laser source is shown with a red curve, indicating the different vertical axes with two arrows. (b) The corrected Sagnac frequency after removing the cavity drift contribution.}
    \label{longterm}
\end{figure}

The Sagnac frequency, $f_s$, which is related to the rotation rate of the PRG, can be obtained from the beat frequency of the heterodyne interferometer:
\begin{equation}
f_s=f_{AOM}-f_{FSR} \label{sag}.
\end{equation}
The beat frequency equals to $f_{AOM}$, which contains the rotation rate $f_s$ and the FSR frequency of the cavity $f_{FSR}$.

\begin{figure}[htbp]
\centering
\includegraphics[width=\linewidth]{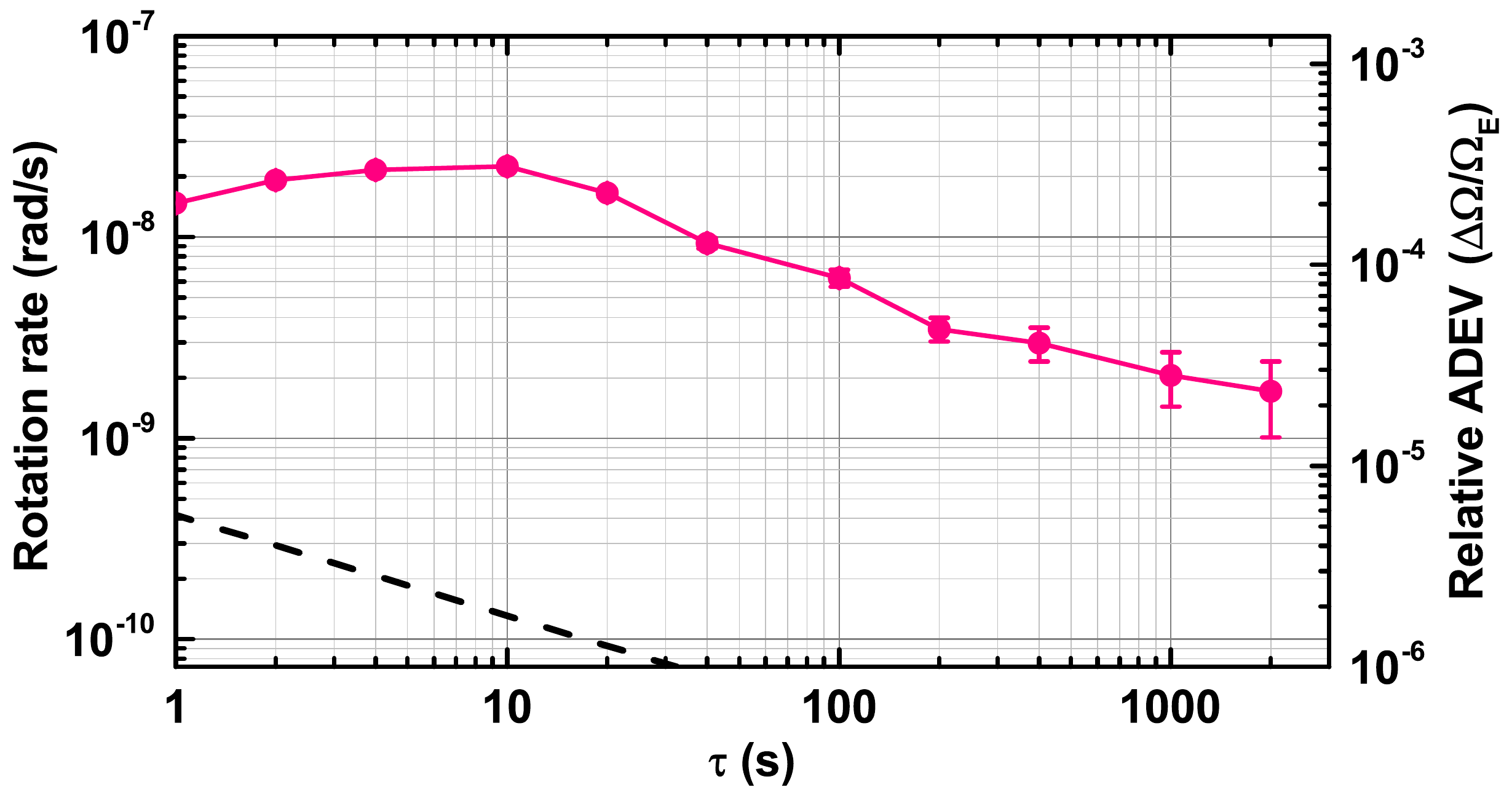}
\caption{Allan deviation of the Sagnac frequency. A relative value to the Earth rotation is shown in the right vertical axis. The black dashed line stands for the shot noise limit.}
\label{allan}
\end{figure}

The ring cavity length change is a common mode noise source if the CW and CCW modes see the same cavity frequency. Since the device operates on adjacent modes with a FSR frequency separation of 75~MHz, the cavity length drift is a major noise source. In our case, the common mode rejection ratio, $CMRR$, is the ratio of the laser frequency to the frequency difference of the two loops, $CMRR=\nu_l/f_{AOM}$. Therefore, we monitor this cavity displacement drift by measuring the resonant PRG laser frequency variation of the primary loop via an ultra-stable laser reference. The ultra-stable laser has a frequency instability of $8\times10^{-16}$ and a drift rate of 3~kHz$/$day \cite{RSI:Zhang:2016, OL:Zeng:2018}. We simultaneously record the Sagnac frequency and the resonant PRG laser frequency of the primary loop with two frequency counters. The results are shown in Fig. \ref{longterm} (a), after the removal of the common linear drift according to Eq. \ref{sag}. It demonstrates a strong correlation between the cavity drift and the Sagnac frequency. We apply a least mean squares adaptive filter to the Sagnac frequency, with the PRG laser frequency as a reference signal. The residual signal after the correction process is shown in Fig. \ref{longterm} (b). The Allan deviation of the corrected rotation rate is depicted in Fig. \ref{allan} and we achieve a resolution of $2\times10^{-9}$~rad/s at an integration time of 1000~s.

\begin{figure}[htbp]
\centering
\includegraphics[width=\linewidth]{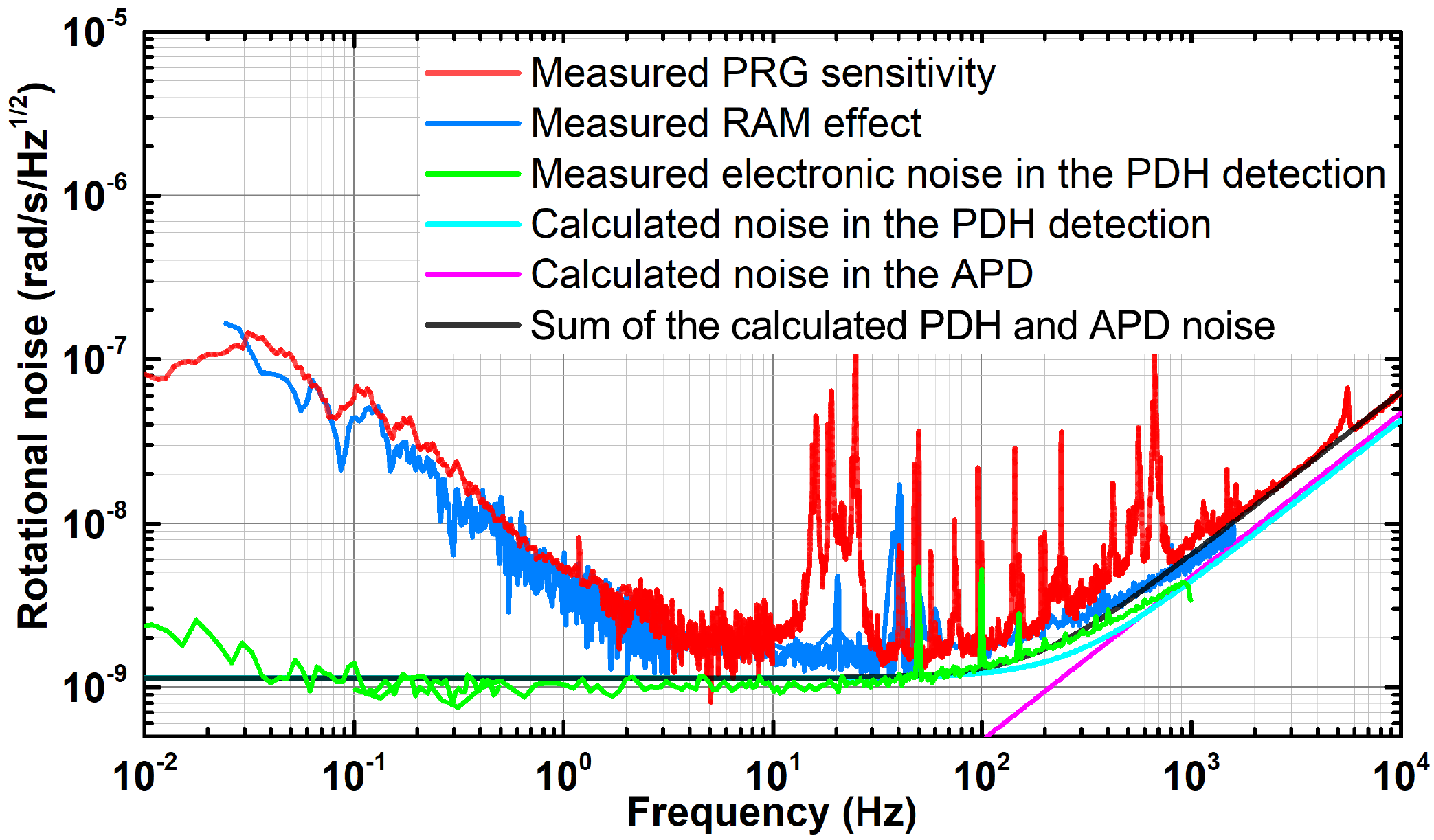}
\caption{The linear power spectral density of the gyroscope output and the estimated noise contributions. The red line is our PRG sensitivity performance, the blue line is the measured RAM effect contribution including electronic noise in the PDH detection, the cyan line is the calculated noise floor in the PDH detection and the pink line is the calculated noise from the APD, and the black line is the sum of the cyan and pink lines.}
\label{psd}
\end{figure}

To further diagnose the performance of our PRG, we measure the frequency noise spectrum of the Sagnac frequency over a large frequency range. The frequency noise above 2~Hz is measured with a phase noise analyzer and the frequency noise below 2~Hz is obtained from a frequency counter. The Sagnac frequency is converted to the rotational noise by the application of the scale factor $\vec{K}$. The result is shown in Fig. \ref{psd} colored in red. The PRG reaches its best sensitivity in the 5-100~Hz frequency range. We have measured the seismic noise with two seismometers simultaneously on and underneath the granite table. From the difference of the two seismic data records, we find that the three peaks around 20~Hz are caused by a tiny resonant motion of the granite platform and its supporting structures. Since a PRG is insensitive to translation, the amplitude of the recorded torsional component amounts to less than 10~nrad.

In order to obtain a theoretical estimation of the noise contributions for the rotational signal shown in the red curve in Fig. \ref{psd}, we have measured the discriminator slopes in the two PDH locking loops as the key parameters. They are obtained by feeding a square wave modulation signal to the error signal of one loop, while the other loop is locked as a reference. The beat frequencies of the two beams result in a square wave shape similar to the modulation signal. As long as the laser frequencies are near the resonance of the cavity, we obtain a linear frequency response. The discriminator slope of both loops are measured to be approximately 1.2~mV/Hz when the power reaching the PDH PD is about 110~$\mu$W. 

We then calculate the contribution of the PD detection noise based on our locking parameters in the PDH locking loop \cite{APB:Shi:2018}. The PDH PD has a noise equivalent power (NEP) of 8~pW$\rm /\sqrt{Hz}$ at the modulated frequencies. The total contribution of the shot noise and electronic noise is calculated to be $\rm1\times 10^{-6}~V/\sqrt{Hz}$, taking into account of the quantum efficiency, the PD response and the cavity transfer function. The rotational noise is dominated by the shot noise and electronic noise on the APD behind the ring cavity at frequencies above 100~Hz. In our case, the NEP of the APD is 2.75~pW$\rm /\sqrt{Hz}$ and the detected power is about 2~$\mu$W. With the measured discriminator slope we can convert all the voltage signals into rotation rate equivalent signals, as shown in Fig. \ref{psd}. The theoretical PDH PD detection noise and the APD detection noise are shown with the cyan and pink lines, respectively. The sum is shown with the black line, which overlaps with our result very well in the high frequency region. Furthermore, we measure the RAM effect contribution including electronic noise in the PDH detection and show it in Fig. \ref{psd} with a blue curve. In this measurement, the RAM effect causes the drift at low frequencies. Since the real RAM contribution can not be directly measured when the PRG is running, we can not remove it at this time. The theoretical and measured results match well, indicating that the current limitations for our PRG are the locking, the RAM effect, and the detection noise.

In conclusion, we have developed a large-scale PRG, which has a sensitivity of $2\times10^{-9}$~rad/s/$\rm \sqrt{Hz}$. It has a comparable sensitivity with the heterolithic active laser gyros with similar sizes \cite{CQG:Virgilio:2010, AO:Beghi:2012}, though it is still one order of magnitude worse than that of the monolithic ring laser C-II \cite{RSI:Schreiber:2013}. To our knowledge, this is the best result among all large scale PRGs \cite{OL:Sanders:1981, CQG:Korth:2016} and we have not yet reached a fundamental limit. This indicates that PRGs have great potential to be a high-resolution Earth rotation sensor. A better performance is expected with better custom-designed photodetectors and a lower shot noise limit. In the future, we plan to use the ultra-stable laser as our light source and increase the side length. The advantage of using an ultra-stable laser is that it can serve as a length standard to stabilize the geometrical scale factor of the PRG, with the obvious advantage of a better long-term stability.

The project is partially supported by the National Key R\&D Program of China (Grant No. 2017YFA0304400), the National Natural Science Foundation of China (Grant No. 91536116, 91336213, and 11774108), and China Postdoctoral Science Foundation (Grant No. 2018M642807).


\end{document}